\def\wgas@ver{2.2}
\def\wgas@pageid{\xdef\@thefnmark{\null}
\@footnotetext{This manuscript was prepared with the
		   AAS WGAS \LaTeX\ macros v\wgas@ver}}
\ifnum\@ptsize<2
\fi
\ps@plain
\def\@tightleading{1.1}
\def\@doubleleading{1.6}
\def\baselinestretch{\@doubleleading}
\def\tightenlines{\def\baselinestretch{\@tightleading}}
\def\received#1{\gdef\@recvdate{#1}} \received{\relax}
\def\accepted#1{\gdef\@accptdate{#1}} \accepted{\relax}
\def\journalid#1#2{\gdef\@jourvol{#1}\gdef\@jourdate{#2}}
\def\articleid#1#2{\gdef\@startpage{#1}\gdef\@finishpage{#2}}
\def\@rcvaccrule{\vrule\@width1.75in\@height0.5pt\@depth0pt}
\def\dates{{\center\small{\it Received:}\space%
\if\@recvdate\relax\@rcvaccrule\else\@recvdate\fi;%
\hspace{1.5em}{\it Accepted:}\space%
\if\@accptdate\relax\@rcvaccrule\else\@accptdate\fi%
\endcenter}}
\let\ltx@abstract=\abstract
\def\abstract{\dates\ltx@abstract}
\def\title#1{{\def\baselinestretch{\@tightleading}
\center\large\bf#1\endcenter}}
\def\author#1{{\topsep 0pt\center\normalsize#1\endcenter}}
\def\affil#1{\vspace*{-2.5ex}{\topsep 0pt\center\small\it#1\endcenter}}

\skip\footins 4ex plus 1ex minus .5ex
\long\def\@footnotetext#1{\insert\footins{
\def\baselinestretch{\@tightleading}\footnotesize
\interlinepenalty\interfootnotelinepenalty
\splittopskip\footnotesep
\splitmaxdepth \dp\strutbox \floatingpenalty \@MM
\hsize\columnwidth \@parboxrestore
\edef\@currentlabel{\csname p@footnote\endcsname\@thefnmark}\@makefntext
{\rule{\z@}{\footnotesep}\ignorespaces
#1\strut}}}
\long\def\@makefntext#1{\noindent\hbox to\z@{\hss$^{\@thefnmark}$}#1}

\def\tablenotetext#1#2{
\@temptokena={\vspace{.5ex}{\noindent\llap{$^{#1}$}#2}\par}
\@temptokenb=\expandafter{\tblnote@list}
\xdef\tblnote@list{\the\@temptokenb\the\@temptokena}}
\def\spewtablenotes{
\ifx\tblnote@list\@empty
\else
\let\@temptokena=\tblnote@list
\gdef\tblnote@list{\@empty}
\vspace{4.5ex}
\footnoterule
\vspace{.5ex}
{\footnotesize\@temptokena}
\fi}
\newtoks\@temptokenb
\def\tblnote@list{}
\def\endtable{\spewtablenotes\end@float}
\@namedef{endtable*}{\spewtablenotes\end@dblfloat}

\def\@xfloat#1[#2]{\ifhmode \@bsphack\@floatpenalty -\@Mii\else
\@floatpenalty-\@Miii\fi\def\@captype{#1}\ifinner
\@parmoderr\@floatpenalty\z@
\else\@next\@currbox\@freelist{\@tempcnta\csname ftype@#1\endcsname
\multiply\@tempcnta\@xxxii\advance\@tempcnta\sixt@@n
\@tfor \@tempa :=#2\do
{\if\@tempa h\advance\@tempcnta \@ne\fi
\if\@tempa t\advance\@tempcnta \tw@\fi
\if\@tempa b\advance\@tempcnta 4\relax\fi
\if\@tempa p\advance\@tempcnta 8\relax\fi
}\global\count\@currbox\@tempcnta}\@fltovf\fi
\global\setbox\@currbox\vbox\bgroup
\def\baselinestretch{\@tightleading}\@normalsize
\boxmaxdepth\z@
\hsize\columnwidth \@parboxrestore}
\def\@keywordtext{Subject headings}
\def\@keyworddelim{---}
\def\keywords#1{\vspace*{-.7ex}
\if@twocolumn\noindent{\small{\it\@keywordtext:\/}\space\@kwds{#1}}
\else{\quote\small{\it\@keywordtext:\/}\space\@kwds{#1}\endquote}
\fi}

\def\@kwds#1{\def\@kwddlm{}\@for\@kwd:=#1\do
{\@kwddlm\def\@kwddlm{\space\@keyworddelim\penalty\@m\space}{\@kwd}}}
\def\section{\@startsection {section}{1}{\z@}{2.3ex plus 1ex minus
.2ex}{1.5ex plus .2ex}{\normalsize\bf}}
\def\subsection{\@startsection{subsection}{2}{\z@}{2ex plus 1ex minus
.2ex}{1ex plus .2ex}{\normalsize\bf}}
\def\subsubsection{\@startsection{subsubsection}{3}{\z@}{2ex plus
1ex minus .2ex}{1ex plus .2ex}{\normalsize\it}}

\def\mathwithsecnums{
\@newctr{equation}[section]
\def\theequation{\hbox{\normalsize\arabic{section}-\arabic{equation}}}}
\def\references{\subsection*{REFERENCES}
\bgroup\parindent=0pt\parskip=\itemsep
\def\refpar{\par\hangindent=3em\hangafter=1}}
\def\endreferences{\refpar\egroup\wgas@pageid}

\def\endthebibliography{\endlist\wgas@pageid}
\def\@biblabel#1{\relax}
\def\@cite#1#2{#1\if@tempswa , #2\fi}
\def\reference{\relax\refpar}
\def\@citex[#1]#2{\if@filesw\immediate\write\@auxout{\string\citation{#2}}\fi
\def\@citea{}\@cite{\@for\@citeb:=#2\do
{\@citea\def\@citea{,\penalty\@m\ }\@ifundefined
{b@\@citeb}{\@warning
{Citation `\@citeb' on page \thepage \space undefined}}%
{\csname b@\@citeb\endcsname}}}{#1}}
\let\jnl@style=\rm
\def\ref@jnl#1{{\jnl@style#1}}
\def\aj{\ref@jnl{AJ}}
\def\araa{\ref@jnl{ARA\&A}}
\def\apj{\ref@jnl{ApJ}}
\def\apjl{\ref@jnl{ApJ}}
\def\apjs{\ref@jnl{ApJS}}
\def\applopt{\ref@jnl{Appl.Optics}}
\def\apss{\ref@jnl{Ap\&SS}}
\def\aap{\ref@jnl{A\&A}}
\def\aapr{\ref@jnl{A\&A~Rev.}}
\def\aaps{\ref@jnl{A\&AS}}
\def\azh{\ref@jnl{AZh}}
\def\baas{\ref@jnl{BAAS}}
\def\jrasc{\ref@jnl{JRASC}}
\def\memras{\ref@jnl{MmRAS}}
\def\mnras{\ref@jnl{MNRAS}}
\def\pra{\ref@jnl{Phys.Rev.A}}
\def\prb{\ref@jnl{Phys.Rev.B}}
\def\prc{\ref@jnl{Phys.Rev.C}}
\def\prd{\ref@jnl{Phys.Rev.D}}
\def\prl{\ref@jnl{Phys.Rev.Lett}}
\def\pasp{\ref@jnl{PASP}}
\def\pasj{\ref@jnl{PASJ}}
\def\qjras{\ref@jnl{QJRAS}}
\def\skytel{\ref@jnl{S\&T}}
\def\solphys{\ref@jnl{Solar~Phys.}}
\def\sovast{\ref@jnl{Soviet~Ast.}}
\def\ssr{\ref@jnl{Space~Sci.Rev.}}
\def\zap{\ref@jnl{ZAp}}

\let\apjlett=\apjl

\def\deg{\hbox{$^\circ$}}

\def\la{\mathrel{\hbox{\rlap{\hbox{\lower4pt\hbox{$\sim$}}}\hbox{$<$}}}}
\def\ga{\mathrel{\hbox{\rlap{\hbox{\lower4pt\hbox{$\sim$}}}\hbox{$>$}}}}

\newcount\lecurrentfam
\def\LaTeX{\lecurrentfam=\the\fam \leavevmode L\raise.42ex
\hbox{$\fam\lecurrentfam\scriptstyle\kern-.3em A$}\kern-.15em\TeX}
\newcommand{\LH}{2}

\newcommand{\SingleSpace}{
  \renewcommand{\LH}{1}
  \def\baselinestretch{\LH}
  \tiny
  \normalsize
}
\newlength{\parindnt}
\newcommand{\PSbox}[3]{\mbox{\rule{0in}{#3}\includegraphics{#1}\hspace{#2}}}
\newcommand{\FigNum}[1]{\unitlength 1pt \begin{picture}(55,10)(-400,35)
			\put(0,0){Figure #1}
			\end{picture}}

\newcommand\sig{$\sigma$}

\newcommand\B{{$B$}}

\newcommand\lt{$<$}
\newcommand\gt{$>$}

\newcommand\etal{et~al.$\!$}

\def\Ref #1 {\lbrack {#1}\rbrack}
\def\vol#1  {{{\bf #1}{\rm,}\ }}
\def\etal   {{et~al.}\ }
\def\apj    {{ApJ{\rm,}\ }}
\def\apjl   {{ApJLett{\rm,}\ }}
\def\apjlett{{ApJLett{\rm,}\ }}
\def\apjs   {{ApJS{\rm,}\ }}
\def\aj     {{AJ{\rm,}\ }}

\def\baas   {{BAAS{\rm,}\ }}
\def\mnras  {{MNRAS{\rm,}\ }}

\def\ssr    {{SSR{\rm,}\ }}
\def\pasj   {{PASJ{\rm,}\ }}
\documentstyle[aasms]{article}
\received{August 25, 1993}
\accepted{October 1, 1993}

\def\dtp{\delta\theta_{pos}}
\def\qla{QL}

\def\cosl{$<\cos~l>$}
\def\sinb{$<\sin^2b> - \frac{1}{3}$}
\def\sig{$\sigma$}
\def\pcs{$\frac{\rm photons}{{\rm cm}^2~s}$}
\def\cpms{$\frac{\rm counts}{1024{\rm ms}}$}
\def\ipll{$I^{1024}_{peak, LL}$}
\def\Ip{$I^{1024}_{peak}$}
\def\IP{\Ip}
\def\B{$B$}
\begin{document}
\SingleSpace

\title{ON THE GALACTIC DISTRIBUTION OF GAMMA-RAY BURSTS}
\author{\sc Robert E. Rutledge and Walter H.~G. Lewin}
\affil{Center for Space Research and Department of Physics \\
Massachusetts Institute of Technology, Room 37-627,
Cambridge, MA 02139, USA }
\begin{abstract}
Recently, \cite{quash93} (hereafter \qla) defined a sub-sample of
Gamma-ray Bursts (GRBs) from the publicly available BATSE database
which shows clumping toward the galactic plane, and they concluded
that all GRBs are galactic in origin.  The selection of these bursts
(duplicated in this work in Sample 1) involved a peak {\it count-rate}
(in \cpms) which is uncorrected for aspect. Using, as limits, the
corresponding peak {\it fluxes } (in \pcs) for the bursts in the
\qla~sample, we find an additional 24 bursts, which we include in a
new sample (Sample 2).  We assert that the peak {\it flux} of a burst
-- the peak {\it count-rate} corrected for detector aspect and energy
response -- is physically more meaningful than peak {\it count-rate},
as used by \qla.  We find that the significance of anisotropy in
Sample 2 is much less than that of Sample 1, which does not support
\qla's interpretation of the anisotropies as being due to a galactic
population.

In addition, to make meaningful statistical statements regarding
isotropy, burst samples must have peak fluxes above a minimum flux,
which is set by the requirement that a burst of certain flux be
detectable from any direction (above the horizon) with respect to GRO,
at any detection threshold at which a burst was observed in that
sample.  We split our Sample 2 into two sub-samples on this basis
(Sample 3 and Sample 4, which have fluxes below and above,
respectively, this minimum flux).  We find that Sample 4 has a
marginal (2.6\sig) deviation from isotropy, which we consider
insufficient to justify the claim that GRBs are galactic in origin.
\end{abstract}


\section{Introduction}
\cite{quash93} (hereafter \qla) define a sample of bursts using two parameters
which
they derive from data in the publicly available BATSE database
(\cite{fishman93}): V (which indicates short-time scale variability)
and \B, which they call burst peak brightness.  Both of these
values are found using the CMAX/CMIN Table available from the BATSE
public catalog.  The CMAX/CMIN Table includes values of
$C^{t}_{\rm max}/C^{t}_{\rm min}$, which is the ratio of peak count-rate
($C^{t}_{\rm max}$, in counts/$t$ ms) in a time bin of duration $t$~ms to the
burst trigger threshold $C^{t}_{\rm min}$ (in counts/$t$ ms) for the time bin
of duration $t$~ms.  It also includes values of $C^{t}_{\rm min}$, so
$C^{t}_{\rm max}$ can be found.  These values are available for $t=64,
256,$ and $1024ms$.

The ``variability'' parameter V (see \cite{lambgraz93}; however, see
also \cite{rutledge93}) is defined:
\begin{equation}
V =   \frac{C^{64}_{\rm max}}{C^{1024}_{\rm max}}
\end{equation}
and the ``brightness'' parameter \B~is simply the maximum peak
counts in a 1024ms bin observed during the burst ($C^{1024}_{\rm max}$).
Thus, the brightness \B~has the units of counts/{1024ms}.  \qla~use
these parameters to select their sample of bursts.  They find that,
for the sample of bursts of log(V)$<-0.8$ (corrected for ``Meegan's
Bias''; see Lamb, Graziani, \& Smith 1993) and $465 \le B \le 1169$,
the bursts are significantly clumped ($\sim$ 4.8\sig~equivalent
Gaussian standard deviations) toward the galactic plane, with a value
\cosl$ = 0.230 \pm 0.078$ and
\sinb $=-0.119\pm0.040$.

In the following discussion and throughout this work, when we mention
peak flux, we mean the maximum 1024~ms averaged flux value, in \pcs,
measured during a burst.

The reader will appreciate that \B~is a count-rate {\it uncorrected for
aspect}; it is not a flux.  This means that, if a given population of
bursts had identical peak fluxes, \B~could vary appreciably depending
on the orientation of the Compton Observatory.  In actuality, we find
that the proportionality of \B~(in counts/1024ms) to $I^{1024}_{peak}$
(found in the publicly available BATSE catalog Flux tables, in \pcs)
can be different by a factor of up to $\sim 3$, with 68\% of the
values to be within 27\% of the mean (see Figure~\ref{fig:alpha}).
Simply put, this means that a burst with a given flux (in \pcs) could
produce values of \B~which differ by up to a factor of 3, and
routinely by a factor of 1.6, due to the orientation of the Compton
Observatory.

The BATSE catalog \B~values come only from the {\it second} most brightly
illuminated BATSE detector, which, at best, could have its normal
directed $\sim 35\deg$ or, at worst, as
much as $70\deg$ from the burst direction.  Since the angular response
of a single BATSE Large Area Detector is similar to $\cos \theta$
(flatter for energies $> 300$ keV; \cite{fishman89}), it is evident
that aspect can cause significantly different values of \B~to be
obtained for bursts of identical peak flux, due only to the orientation of
the Compton Observatory.

Clearly, use of \B~as a criterion for a burst population study opens
up the possibility of a directionality bias.  This can be overcome by
setting a lower limit on the peak flux (\ipll) on the bursts included,
such that at all times it is possible to detect a burst with peak flux
\ipll~from any unocculted direction.

\ipll~can be roughly approximated using data from the public
BATSE database.  Assume a burst with peak flux \IP~is incident
on the BATSE detectors from a direction with transmission efficiency
$\alpha$, approximated by:
\begin{equation}
\label{eq:alphadef}
\alpha~I^{1024}_{\rm peak} = B = \frac{C^{1024}_{\rm max}}{C^{1024}_{\rm
min}}~C^{1024}_{\rm min}
\end{equation}
where \B~is in $\frac{counts}{1024 ms}$, \IP~is in \pcs, and we've
included the time proportionality (1~sec/1024ms) in $\alpha$. If
\ipll~is to be minimally detectable (i.e.,
$\frac{C^{1024}_{\rm max}}{C^{1024}_{\rm min}}=1.0$), from any
unocculted angle (that is,
at the lowest possible $\alpha$) at any time (when the detection
threshold $C^{1024}_{\rm min}$ is highest), then we find:

\begin{equation}
I^{1024}_{peak, LL} = MAX(C^{1024}_{\rm min})\frac{1}{MIN(\alpha)}
\end{equation}

where $MAX(C^{1024}_{\rm min})$ is the maximum of all values
$C^{1024}_{\rm min}$ for bursts considered and $MIN(\alpha)$ is the
minimum of all values of $\alpha$.  We here make the assumption that
the 207 bursts in the public database for which $\alpha$ values can be
found have more or less explored the unocculted aspect-space of BATSE.  Using
data from the database, we find $MAX(C^{1024}_{\rm min})=346$,
$1/MIN(\alpha)=0.001948$, (where we have ignored one exceptionally
high value of 1/$\alpha$, from burst number 1346, for which the
$(C^{1024})_{max}$ value listed in the BATSE public database is
probably erroneous; see \cite{lambgraz93}), and therefore our
\ipll=0.674 \pcs.  There are 104 bursts in the public database above
this flux limit (i.e. which could have been detected from any
unocculted direction at any time during the BATSE observation period) for
which $\alpha$ can be found.  If each has a systematic positional
error box of $4\deg$ which does not overlap with another in
aspect-space, then the aspect space would be 101\% filled.  The
assumption that the aspect space is completely explored is therefore,
perhaps, not unreasonable.

It is possible to limit the sample of our bursts to those detected at
times with lower background, and thus find a lower $MAX(C^{1024}_{\rm min})$,
which would lower the \ipll~as well.  However, to calculate the
significance of any such result, we require knowledge of the Sky
Exposure during which $C^{1024}_{\rm min} \le MAX(C^{1024}_{\rm min})$, and as
the public database currently lists only the complete Sky Exposure, we
must take the highest $C^{1024}_{\rm min}$ observed, producing the
subsequently higher \ipll.

\section{Analysis}
We defined four burst samples, described below and shown in
Figure~\ref{fig:samples} as a function of \B~and burst peak flux.  The
two vertical lines in this figure are the $C^{1024}_{\rm max}$ limits
used by \qla.  The two solid horizontal lines are the {\it second}
highest and {\it second} lowest peak fluxes in the sample defined by
\qla, and the broken horizontal line is \ipll.  We use these fluxes in
the definitions of Samples 2, 3, \& 4:

{\bf Sample 1.} Sample 1 (shown in Figure~\ref{fig:samples}a) is
identical to the burst sample used by \qla.  Using the \B~and V values
kindly provided (Carlo Graziani, private communication),  bursts were
selected which had 465~$\le$~B~$\le$~1169 \cpms, and log(V)$\le$-0.8.
There are 55 bursts which meet these criteria.

{\bf Sample 2.} Sample 2 (Figure~\ref{fig:samples}b) includes bursts
with peak fluxes $0.396\le I_{peak} \le 1.296~$\pcs, with $\dtp \le
10\deg.77$, and with log(V)$\le$-0.8. $\dtp$ is the total positional
error box.  We find 79 bursts meeting these criteria.

{\bf Sample 3.} Sample 3 (Figure~\ref{fig:samples}c) includes bursts
with peak fluxes $0.396\le I_{peak} \le 0.674~$\pcs, with $\dtp \le
10\deg.77$, and with log(V)$\le$-0.8.  Sample 3 is a subset of Sample 2.
We find 40 bursts meeting these criteria.

{\bf Sample 4.} Sample 4 (Figure~\ref{fig:samples}d) includes bursts
with peak fluxes $0.674\le I_{peak} \le 1.296~$\pcs, with $\dtp \le 10\deg.77$,
and with log(V)$\le$-0.8.  Here we have used the \ipll~found above as
our lower flux limit.  We find 39 bursts meeting this criteria. Sample
4 is a subset of Sample 2, and is complimentary to Sample 3.

Figure~\ref{fig:aitoff} shows positional mappings in Galactic
coordinates of each of these four samples.

Using the Galactic positions of each burst, we produce values of
\cosl~and \sinb~for each of the four samples.

\section{Results}
The results of this analysis are shown in Table~\ref{tab:results},
along with the result of the analysis performed by \qla.  Due to
unequal sky coverage, the standard deviations quoted in the \cosl~and
\sinb~columns are not Gaussian, although because the sky coverage is
not enormously uneven (i.e. varies by $\sim$ 40\%), the standard
deviations can be taken to be very {\it roughly} of Gaussian
significance.

Sample 1, selected using identical criteria to \qla~, produces a value
of \cosl~and which is different from that found by
\qla~although it is identical in significance ($+0.272\pm 0.093$
vs.$+0.230\pm 0.078$; both 2.9\sig).  The values of
\cosl~and its uncertainty are both larger than those found
by \qla~by a factor of $\simeq1.18$, which is roughly $\frac{65}{55}$.
The value of \sinb~is identical to that found by \qla,
although the significance we find is greater than that found by
\qla($-0.119 \pm 0.030$, vs. $\pm 0.040$, which is 4.0\sig~vs. 3.0\sig).

Sample 2, selected using the more physically meaningful {\it flux} limits
similar
to those which exist in Sample 1, produces values of
\cosl~and \sinb~with considerably lower significance than Sample 1
(1.8\sig~and 2.4\sig, respectively).

Sample 3, composed of bursts from Sample 2 which have
\IP \lt  \ipll,  produces a stronger dipole (2.1\sig) than Sample 2, but no
significant quadrupole (0.8\sig).

 Sample 4, composed of bursts from Sample 2 which have
\IP \gt \ipll~produces a stronger quadrupole (3.0\sig) than Sample 2, but no
significant dipole (0.5\sig).

\qla~quantify the significance
of their result through Monte Carlo simulations to find the
probability of producing their observed dipole and quadrupole moments
from an isotropic distribution, correcting for sky coverage.
However, the Sky Exposure
Table of the BATSE public database states that the Sky Exposure
values apply ``to bursts intense enough to trigger the instrument from
any direction not occulted by the earth,'' (that is, for bursts with
$I^{t}_{peak} \ge I^{t}_{peak, LL}$, t=64, 256, or 1024ms).  We find
that approximately one third of the bursts in the
\qla~sample do not meet this criterion; had the
Compton Observatory been pointed in a different direction, or had they
occurred during periods of higher background, these bursts may not
have been detected by BATSE.  Because \qla~used the Sky Exposure Table
to calculate the probability of detecting bursts which the Sky
Exposure Table excludes, the probabilities \qla~derives using the Sky
Exposure Table may not apply. The same is true for our Samples 1, 2 \&
3; the probabilities for the deviations from isotropy in these samples
cannot be calculated in a straightforward manner.

We performed Monte Carlo simulations (see Appendix A) to find the
significance of the deviations from isotropy found in Sample 4, which is
composed
entirely of bursts with \IP \gt \ipll.  We note that
none of the 39 bursts from Sample 4 were ``overwrites'' -- bursts
which triggered the BATSE detectors prior to the readout period of an
earlier, weaker burst; these cannot be used in conjunction with the
Sky Exposure Table, as they occur during detector ``dead-time''.

Taking into account sky coverage, we find a probability of
producing equal or greater \cosl~and \sinb~terms as in Sample 4 to be
0.90\% (=2.6\sig, see Appendix A).

\section{Discussion and Conclusions}

We find that the sample of BATSE bursts used by \qla~does not
constitute a flux-selected sample, but is based on the (physically
irrelevant) pointing of the GRO satellite.

We also find that a sample of BATSE detected bursts will be
directionally biased if steps are not taken to insure that all bursts
within a defined sample were detectable at the highest detection
threshold limit (i.e. for the highest value of $C^{t}_{\rm min}$) used
within the sample {\it and} for all aspect angles.  We point out that
with a more exact knowledge of the angular response,
\ipll~could be more precisely defined than we have done so here.

In going from a ``brightness''-selected sample (Sample 1) to a
flux-selected sample (Sample 2) within the flux limits of the
``brightness'' sample, we find that the non-Gaussian significances of
\cosl~and \sinb~drop considerably (from 2.9\sig~and 4.0\sig, to
1.8\sig~and 2.4\sig).  This does not support \qla's interpretation of
an anisotropy in the GRB distribution as due to a galactic origin.
Because the flux limits used in this sample include bursts with fluxes
well below the flux completion limits of the BATSE database, it is not
straightforward to estimate significances of this measurement in the
absence of directionally specific (i.e., RA and dec.)  flux-detection
efficiencies.

While the probability of producing the observed anisotropies from a
purely isotropic distribution of bursts on the sky is small (0.9\%),
we feel that it is not small enough to justify the claim that GRBs are
of galactic origin.

\section*{Acknowledgements}
The authors thank Chyrssa Kouveliotou and Chip Meegan for their
comments on this manuscript prior to submission, and Carlo Graziani
for providing the \B~and corrected V values for each burst. This work
was supported through NASA grant \#NAGW-3234.

{\it Note added in proof: It was announced at the Hunstville Gamma Ray
Burst Workshop 1993 by C. Meegan that bursts selected in the
brightness range 490 $<$ B $<$ 1250 from 480 bursts observed by BATSE not
included in the first BATSE GRB catalog have no significant deviation
from an angularly isotropic distribution.  This is in support of our
conclusions.

We have contacted C. Kouveliotou and G. Pendleton regarding
uncertainties in the BATSE flux values in the first BATSE catalog.
The upper limits on systematic errors in the peak fluxes are 10-15 per
cent, which is much lower than the known systematic errors, due to
angular response, in peak counts as used by as QL.}

\section*{ Appendix A: Monte Carlo Simulations}
\appendix

We outline the Monte Carlo simulations used to determine the
significance of the deviations from isotropy.

We have a burst sample of N bursts with observed values of
\cosl$_{obs}$ and (\sinb)$_{obs}$.

We perform a simulation, placing N bursts randomly on the sky in RA
and Declination, $\alpha_i$ and $\delta_i$ with angles (for the
$i^{th}$ burst):
\begin{eqnarray}
	\alpha_i & = & 2\pi X   \nonumber \\
	\delta_i & = & \arccos(1 - 2Y) - \frac{\pi}{2}  \nonumber
\end{eqnarray}
where $X$ and $Y$ are random uniform deviates between 0 and 1,
inclusive (using a random number generator from \cite{press}), and $ 1
\le i \le N$.  We then look up the total exposure time ($E_i$) for $\delta_i$
in the BATSE Sky Exposure Table.  We
transform from RA/DEC to Galactic co-ordinates ($\alpha_i$, $\delta_i$
to $b_i$, $l_i$), and find the values for the simulated sample of
bursts:
\begin{eqnarray}
	<\cos~l> & = & \frac{ \sum_{i=1}^{N}
E_i~\cos(l_i)}{\sum_{i=1}^{N} E_i}   \nonumber  \\
	<\sin^2~b> & = & \frac{\sum_{i=1}^{N} E_i~\sin^2(b_i)}{\sum_{i=1}^{N} E_i}
\nonumber
\end{eqnarray}
We perform this simulation of N bursts continually, keeping a tally of
$N_{trials}$ (the number of N-burst simulations) and $N_{success}$,
which  is incremented when the following criteria are met:
\begin{eqnarray}
	|<\cos~l>| &\ge& |<\cos~l>_{obs}|  \nonumber \\
	<\sin^2~b> &\le& <\sin^2~b>_{obs}  \nonumber
\end{eqnarray}
since we are interested only in matching or exceeding the magnitude
(not the sign) of the dipole, and since we must also match or exceed
the quadrupole.   When $N_{success}$ is large (\gt 3000), we stop the
simulations, and take the probability of matching or exceeding the
observed Galactic dipole and quadrupole
\begin{eqnarray}
 \nonumber
	Q = \frac{N_{success}}{N_{trials}}
\end{eqnarray}
We then look up the corresponding tabulated Gaussian standard
deviation value in Bevington (1969) which encompasses $1-Q$ of all
values. For instance, if $Q=0.32$, then 1 Gaussian standard deviation
encompasses 1-0.32=0.68 of all measured values.   We call this this an
``equivalent Gaussian standard deviation'' of 1\sig.

\clearpage

\begin{table}[h]
\begin{center}

\caption{Results \label{tab:results}}
\vspace{0.1cm}
\begin{tabular}{lccccc} \hline \hline
		& Flux Limits		& Number   	&  					&   			     &   	\\
Sample 		& (\pcs)		&  of Bursts 	& \cosl$^a$				& \sinb$^a$ 		     &  Q   	\\
\hline
\qla~Result	&     -			&   55		&   $+0.230\pm 0.078$~(2.9\sig)		& $-0.119\pm
0.040$~(3.0\sig)&  -$^b$	\\
1	&     -			&   55		&   $+0.272\pm 0.093$~(2.9\sig) 	& $-0.119\pm
0.030$~(4.0\sig)&  -$^b$	\\
2 	&0.396$\le$\IP$\le$1.296&   79		&	$+0.141\pm 0.077$~(1.8\sig)	& $-0.070\pm
0.029$~(2.4\sig)&  -$^b$	\\
3 	&0.396$\le$\IP$\le$0.674&   40		&	$+0.233\pm 0.111$~(2.1\sig)	& $-0.035\pm
0.047$~(0.8\sig)&  -$^b$ 	\\
4 	&0.674$\le$\IP$\le$1.296&   39		&	$+0.048\pm 0.104$~(0.5\sig)	& $-0.106\pm
0.035$~(3.0\sig)&  0.0090	\\ \hline
\multicolumn{6}{l}{$^a$ Standard deviations given parenthetically are only {\it
roughly} Gaussian (see text)} \\
\multicolumn{6}{l}{$^b$ Probabilities not calculated (see text)} \\
\end{tabular}
\end{center}
\end{table}
%

\newpage

\begin{figure}
\caption{ \label{fig:alpha}
Ratio of B to $I^{1024}_{peak}$ for bursts in the BATSE public database.
This ratio is identical to the $\alpha$ we use in Equation~2.}
\end{figure}

\begin{figure}
\caption{ \label{fig:samples}
Peak Flux (in photons/cm$^2$/sec) (from the flux tables of the BATSE
public database) vs. brightness B(= $C^{1024}_{\rm max}/ C^{1024}_{\rm min} *
C^{1024}_{min}$, both from the CMAX.CMIN table of the BATSE public
database).  The two vertical lines are the
$C^{1024}_{\rm max}$ limits used by
\qla.  The two horizontal bars are the {\it second} highest peak
fluxes in the sample defined by \qla.  (a) Bursts selected into Sample
1, following the criterion of \qla.  (b) Sample 2 bursts. (c) Sample 3
bursts. Sample 3 is a subset of Sample 2.  (d) Sample 4
bursts.  Sample 4 is a subset of Sample 2, and is complimentary to
Sample 3. }
\end{figure}


\begin{figure}
\caption{ \label{fig:aitoff}
Projection mappings of four samples of GRBs for the present work, in
galactic coordinates.  The sample number is indicated above each
panel. }
\end{figure}

\clearpage

\clearpage
\pagestyle{empty}
\begin{figure}
\PSbox{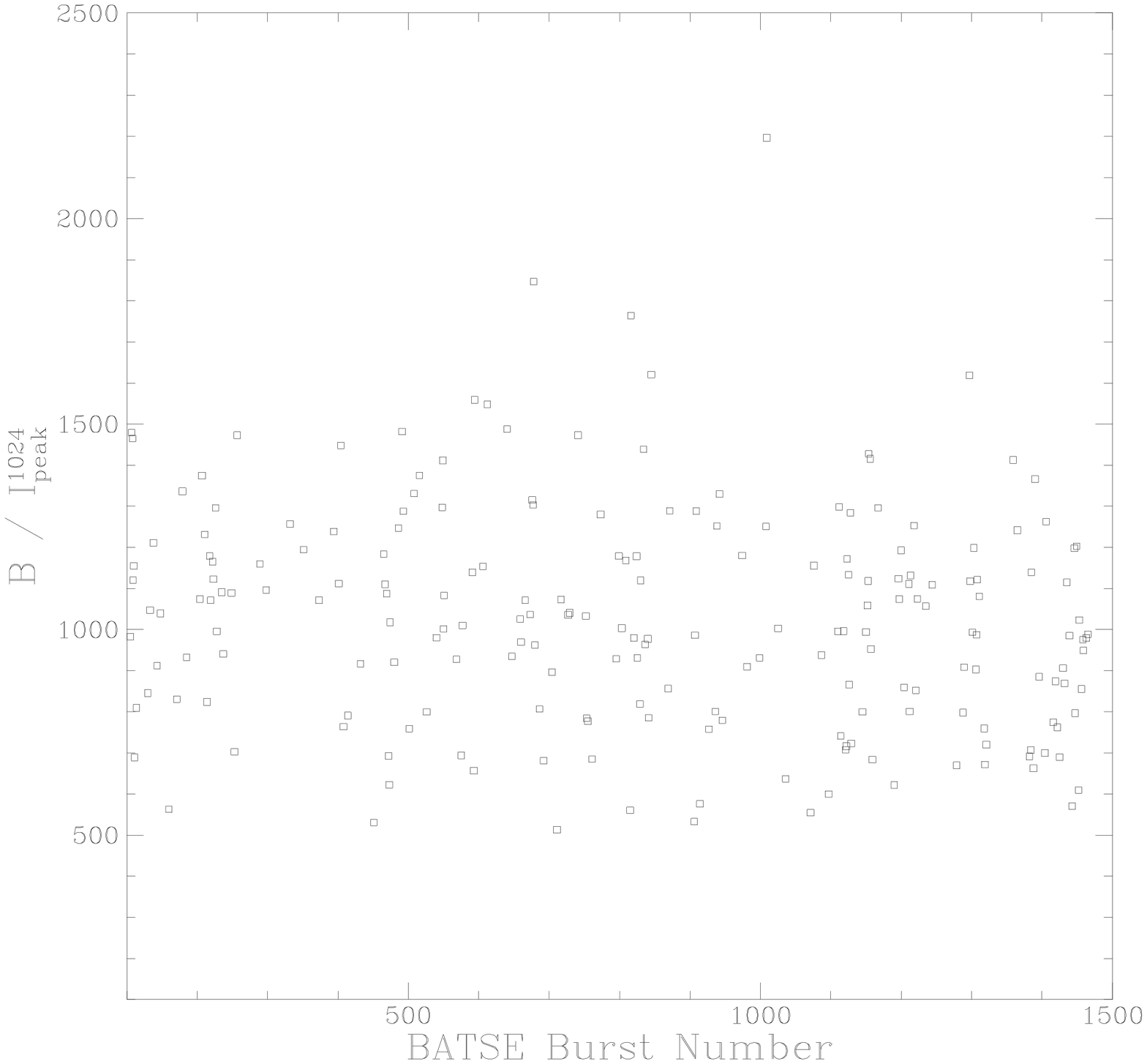 hoffset=-85 voffset=-150}{14.7cm}{17.78cm}
\FigNum{\ref{fig:alpha}}
\end{figure}

\clearpage
\pagestyle{empty}
\begin{figure}
\PSbox{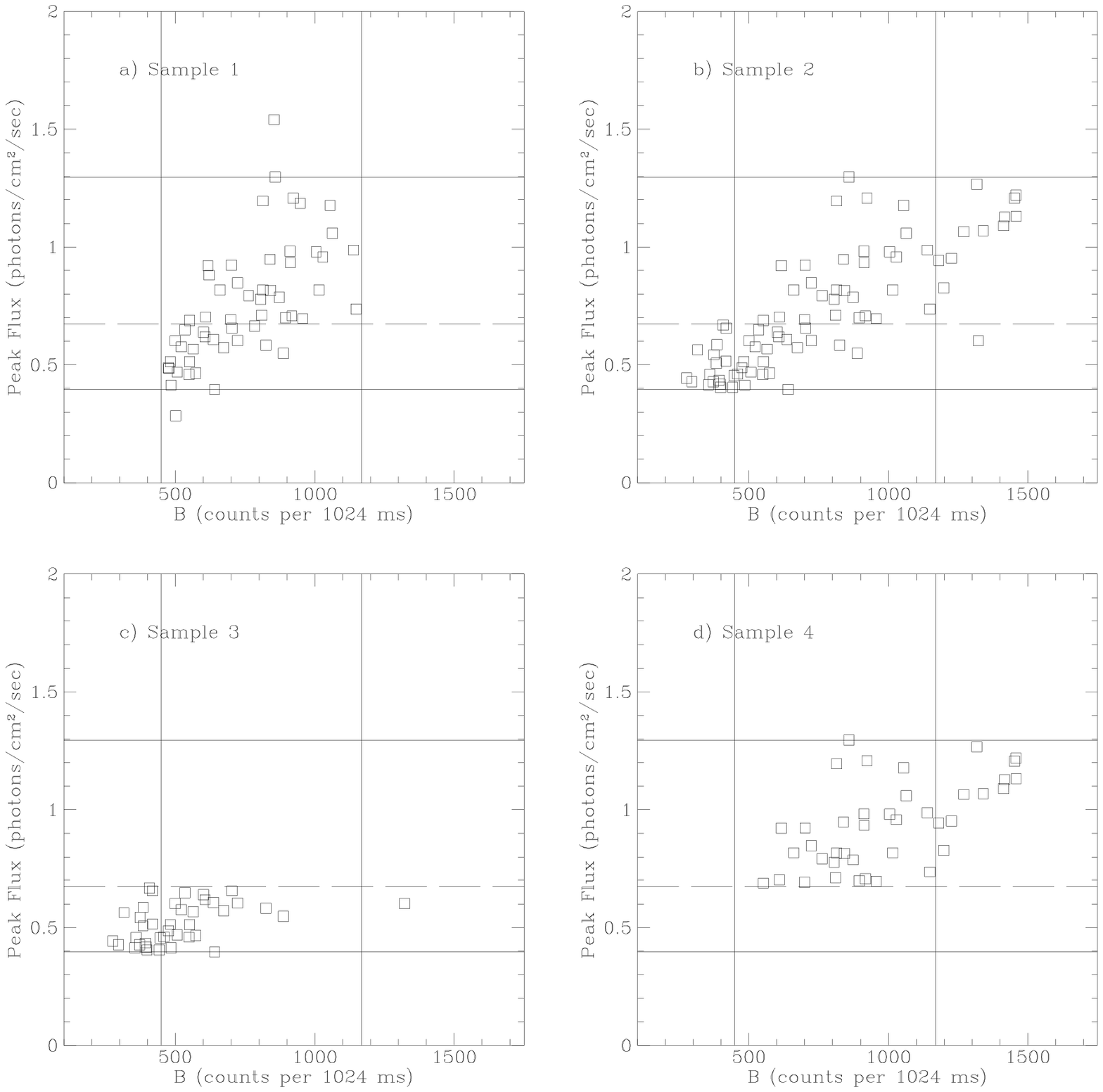 hoffset=-85 voffset=-150}{14.7cm}{17.78cm}
\FigNum{\ref{fig:samples}}
\end{figure}

\clearpage
\pagestyle{empty}
\begin{figure}
\PSbox{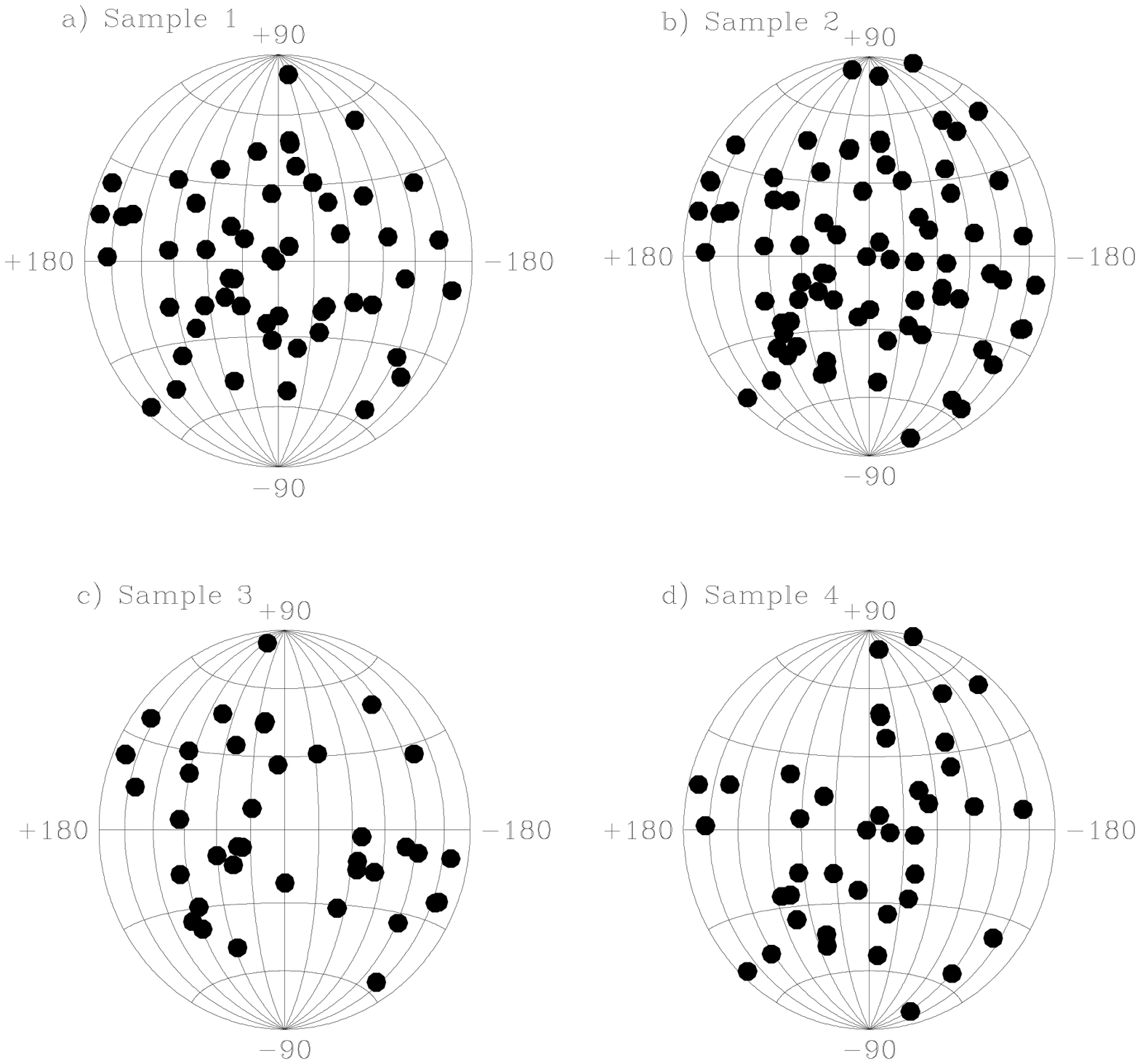 hoffset=-85 voffset=-150}{14.7cm}{17.78cm}
\FigNum{\ref{fig:aitoff}}
\end{figure}

\end{document}